\documentclass[12pt]{article}
\usepackage{epsfig}
\begin{document}

\begin{center}\begin{large}{\bf Experimental Chiral Dynamics:
New Opportunities with Polarized Internal Targets and Almost-Real 
Photon Tagging}\end{large}\\
\vspace{1.5mm}
A.M. Bernstein and M.M. Pavan \\
Physics Department and Laboratory for Nuclear Science\\
 M.I.T., Cambridge, MA, U.S.A.
\end{center}

\begin{abstract}
  Experiments on pion (Goldstone Boson) photoproduction from the nucleon
  tests the ability to make QCD predictions at confinement scale energies.
  Experiments with both polarized beams and targets have the potential
  sensitivity to demonstrate the dynamic isospin breaking effects of the up
  and down quark mass differences, whereas experiments on Compton
  scattering from the nucleon will incisively probe its chiral structure by
  measuring all of the spin dependent amplitudes.  These and other types of
  photo--induced measurements on nuclei could be possible at BLAST
  with the addition of an almost--real photon tagging system and a forward
  low   energy recoil ion hodoscope.

\end{abstract}

\section{Confinement Scale QCD and Chiral Dynamics}
\label{sec:intro}

At low energies the interaction between quarks and gluons is extremely
strong and leads to confinement, where approximate QCD solutions can be
obtained by an effective field theory known as chiral perturbation theory
(ChPT) or Chiral Dynamics\cite{book,ChPT,ChPT2,work1}.  This is based on
the chiral symmetry present in the QCD Lagrangian in the limit of massless
light quarks, but which is broken in the ground state of matter. In such a
situation, Goldstone's theorem states that there are massless, pseudoscalar
Bosons whose interactions with other hadrons vanish at zero
momentum\cite{book,Goldstone,L}. In the case of m$_{u}$=m$_{d}$=0, there
are three Goldstone Bosons which are identified as the pion triplet. The
relatively weak interactions of Goldstone Bosons at low energies invites a
perturbation scheme based on chiral symmetry and hadronic degrees of
freedom.

In the real world, the light quark masses are nonzero, but small
\cite{W1,GL1}. Therefore, for strong interaction theory to have predictive
power, calculations must be performed taking the deviations from the pure
Goldstone theorem into account. As an example, the s wave scattering
length, $a$, vanishes for a Goldstone Boson scattering from any hadron in
the low energy limit. However for a physical meson with finite mass
($\pi,\eta,K$) one would intuitively expect $a \simeq 1/\Lambda_{x}$ 
(see contribution of A.B. in Ref.\cite{work1}) where
$\Lambda_{x}$ is the chiral symmetry breaking scale ($\simeq
4\pi f_{\pi}\simeq 1 GeV$ for pions, where $f_{\pi}= 92.4 MeV$ is the
decay 
constant). This intuitive expectation is supported by the original
calculation of Weinberg \cite{W2} in which the scattering lengths of pions
from any hadrons were first obtained by current algebra techniques (a
precursor of ChPT). The order of magnitude of the s wave scattering lengths
is\cite{W2}:
\begin{equation}
 a_{o} = \frac{m_{\pi}}{4 \pi f_{\pi}^{2}}=\frac{m_{\pi}}{\Lambda_{x}
f_{\pi} }\simeq \frac{1.5}{\Lambda_{x}}
\end{equation}
One observes that $a_{o}\rightarrow 0$ when $m_{\pi} \rightarrow 0$ 
(the chiral limit) and also that $a_{o} \simeq 1/\Lambda_{x}$.

Similarly, one would expect the production and decay amplitudes of
Goldstone Bosons to vanish in the chiral limit.  Some examples, which can
be obtained from ChPT calculations \cite{ChPT,ChPT2}, are the threshold
electric dipole amplitude, $E_{0+}^{\gamma N \rightarrow \pi^{0}N}$ for s
wave photo-pion production, the $\Sigma$ term of $\pi N $ scattering , the
isospin breaking $\eta \rightarrow 3 \pi$ decay, and the form factors for
$K_{l4}$ decays
In a similar vein, there are some observables that
diverge in the chiral limit, such as the charge radii and polarizabilities
of nucleons and pions \cite{ChPT,ChPT2}. In this case, the physical
interpretation is that the meson cloud extends beyond the hadron and in the
chiral limit extends to infinity.

Pion-hadron scattering and the amplitudes given above are examples of
quantities that either vanish or blow up in the chiral limit. In the real
world, where the light quark masses are non-zero, chiral symmetry is
explicitly broken and these quantities are finite and non-zero. Their
precise (nonzero and finite) values are measures of explicit chiral
symmetry breaking, and therefore a theoretical challenge to calculate them.
Quantities which either vanish or diverge in the chiral limit point to an
experimental opportunity to perform precise experiments, not only to check
ChPT calculations, but also as fundamental quantities which must be
predicted by any theory of the strong interaction. Experimental Chiral
Dynamics is the study of the properties, production and decay amplitudes,
and low energy interactions of the almost Goldstone Bosons ($\pi,\eta,K$)
with themselves and with other hadrons.

The main purpose of this contribution is to point out new and exciting
experimental possibilities in chiral dynamics which arise by having an
experimental apparatus with thin, pure, polarized, targets in an electron
storage ring with an intense, polarized, electron beam of $\simeq$ 1 GeV.
We will introduce the possibility of a new very small angle electron
scattering/almost--real photon tagging
facility (SMASH) which will utilize the polarized, internal targets being
built for BLAST. As important examples we shall discuss the threshold
$\vec{\gamma} p \rightarrow \pi^{0} p$ and polarized Compton scattering,
$\vec{\gamma} \vec{p} \rightarrow \gamma p$, reactions. Furthermore,
such a  facility would be capable of measuring all
photo-hadron processes with polarized photons on polarized and unpolarized
internal targets, e.g. protons, deuterons, and $^{3}He$. In particular, the
coherent $\vec{\gamma} 
\vec{D} \rightarrow \pi^{0} D$ reaction can be accessed from threshold
through the $\Delta$ region and could produce important new results on the
$\vec{\gamma} \vec{n} \rightarrow \pi^{0} n$ amplitude. 
Some of these experiments require soft recoil ion
detection near the internal target.
We will also
point out the possibility of using the BLAST detector itself in addition to 
the internal target to investigate other timely physics issues e.g. 
to study the quadrupole components
in the $\vec{\gamma} \vec{p} \rightarrow \Delta$ transition.

\subsection{Photopion Reactions and Light Quark Dynamics}

Near threshold pion photoproduction is an excellent example of confinement
scale QCD physics, where considerable theoretical and experimental progress
has been made in the past few years.  Starting with the availability of CW
electron 
beams, the $\gamma p \rightarrow \pi^{0}p$ reaction has been performed with
high quality tagged photon beams at Mainz\cite{Mainz} and
Saskatoon\cite{Sask}.  At the same time, ChPT calculations have been
performed which have advanced our understanding of this reaction\cite{BKM}.
For the sake of brevity only the photoproduction experiments will be
discussed here.

ChPT is an effective field theory which uses the observed hadrons rather
then the quarks and gluons as the degrees of
freedom\cite{book,ChPT,ChPT2,work1}. The effective Lagrangians are
organized into a series of increasing powers of the momenta,
$(p/\Lambda_{x})^{n}$, where $\Lambda_{x} \simeq 4 \pi f_{\pi} \simeq 1
GeV$ is the chiral scale parameter. The introduction of a higher order
Lagrangian introduces low energy constants which are required to
renormalize the infinities order by order.  At the present time these low
energy constants must be determined empirically by fits to data or
estimated by the principle of resonance saturation. In principle they can
be obtained from the QCD Lagrangian by integrating out the high energy
degrees of freedom, e.g.  by lattice gauge theory. The importance of ChPT
is that it is an effective {\it theory} based on QCD, i.e. at each order in
the momentum expansion, the diagrams that must be calculated are specified
and not left to individual discretion as they are in model calculations.

At the present time one loop ChPT calculations for electromagnetic pion
production have been carried out to $O(p^{4})$ \cite{BKM}.  The presence of
the counterterms implies 3 low energy constants in photoproduction and 5 in
electroproduction: two in the transverse s wave multipole $E_{0+}$, one in
the p wave transverse multipoles, and two in the longitudinal s wave
multipole $L_{0+}$ (for electroproduction). Currently, these are
determined by a fit to the data and also estimated by the principle of
resonance saturation. The two approaches are found to agree \cite{BKM},
indicating that the values are understood. It should also be noted that
$\pi^{0}$ photo and electroproduction from the neutron can be predicted
without any additional parameters. Thus measurements of the neutron
production amplitudes will provide a stringent test of ChPT calculations.

One important advantage that ChPT has brought to the study of
electromagnetic meson production is the systematic ordering of the
diagrams. In particular, pion rescattering in the final state (1 loop
diagram) is the crucial ingredient of the near threshold energy dependence.
The largest contribution comes from the production of charged pions in the
intermediate state. Since the ratio of the electric dipole amplitudes for
the neutral and charged pion channels $R= E_{0+}^{\gamma p \rightarrow
  \pi^{+}n} /E_{0+}^{\gamma p \rightarrow \pi^{0}p}\simeq -20$, the two
step $\gamma p \rightarrow \pi^{+}n\rightarrow \pi^{0}p$ reaction is as
strong as the direct $\gamma p \rightarrow \pi^{0}p$ path.  Combined with
the separation of the $\pi^{0}p$ and $\pi^{+}n$ thresholds this leads to a
unitary cusp in the $\gamma p \rightarrow \pi^{0}p$ reaction.

\begin{figure}[ht]
  \begin{center}
\epsfig{file=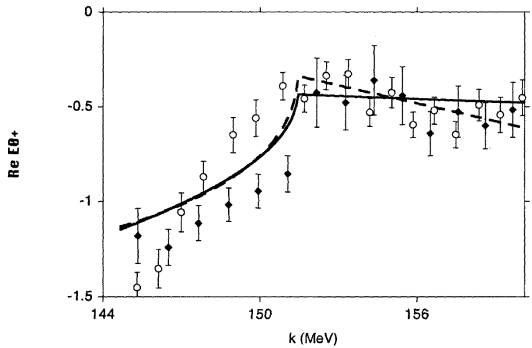, width=0.45\textwidth}
    \caption{ \footnotesize
$Re E_{0+}$ (in units of $10^{-3}/m_{\pi}$) for the 
$\gamma p \rightarrow \pi^{0}p$ reaction versus photon energy k. 
The dashed dot curve is the ChPT fit \protect\cite{ChPT} and the 
solid curve is the unitary fit Eq. \protect\ref{eqn:e0+_unitary} to the
Mainz \protect\cite{Mainz} data (open circles). The Saskatoon data
\protect\cite{Sask} is also shown (filled circles).
 \normalsize}
    \label{fig:ree0+}
  \end{center}
\end{figure}

The simplest example to understand the occurrence of the unitary cusp
in the $\gamma^{*}p \rightarrow \pi^{0}p$ reaction (where $\gamma^{*}$ is a
real or virtual photon) is to use the 3 channel S matrix for the open
channels ($\gamma^{*} p, \pi^{0}p$, $\pi^{+}n$ )\cite{AB}.  Applying the
constraints of unitary and time reversal invariance, one is led to the
coupled channel result for the s wave amplitude $E_{0+}^{\gamma p
  \rightarrow \pi^{0}p}$:
\begin{equation}
 E_{0+}^{\gamma p \rightarrow \pi^{0}p}(k) 
                             = e^{i\delta_{0}}  [A(k) + i \beta q_{+} ] 
\label{eqn:e0+_unitary}
\end{equation}     
where $\delta_{0}$ is the s wave $\pi^{0}p$ phase shift (predicted to be
very small), A(k) is a smooth function of the photon energy k, $\beta=
E_{0+}^{\gamma p \rightarrow \pi^{+}n} \cdot
a^{cex}_{\pi^{+}n\rightarrow\pi^{0}p}$ is the cusp parameter, and $q_{+}$
is the $\pi^{+}$ CMS momentum which is continued below the $\pi^{+}n$
threshold as $i\mid q_{+}\mid$. The cusp function $\beta q_{+}$ contributes
to the real (imaginary) part of $E_{0+}$ below (above) the $\pi^{+}n$
threshold.

The results for the real part of the s wave electric dipole amplitude
$E_{0+}^{\gamma p \rightarrow \pi^{0}p}$ are presented in Fig.
\ref{fig:ree0+}. The rapid energy dependence between the $\pi^{0}p$ and
$\pi^{+}n$ thresholds at 144.7 and 151.4 MeV due to the unitary cusp can be
seen. Above the $\pi^{+}n$ threshold the energy dependence is much less
rapid. This is in approximate agreement with the predictions of the unitary
cusp (Eq.\ref{eqn:e0+_unitary},Ref.\cite{AB}) and ChPT\cite{BKM}. It should
be noted that the errors in the  
data shown in Fig.\ref{fig:ree0+} are statistical only so that the
disagreement between the data sets  
is not serious. To complete the observation of the unitary cusp and to
precisely measure the value of $\beta$, $Im(E_{0+})$ must be measured.
In particular, experiments with polarized targets can measure the predicted 
rapid rise in $Im (E_{0+}^{\gamma p \rightarrow \pi^{0}p})$ above the
$\pi^{+}n$ threshold. The SMASH facility
outlined here would provide such important data.

The unitary constraints to the multipole amplitudes show the importance of
the final state $\pi N$ scattering and charge exchange on the $\gamma^{*} p
\rightarrow \pi^{0}p$ reaction. It raises the possibility of measuring
$a_{\pi^{0}p}$ and $a^{cex}_{\pi^{+}n \rightarrow \pi^{0} p}$ by
measurements of $Im(E_{0+}^{\gamma p \rightarrow \pi^{0}p})$ both below and
above the $\pi^{+}n$ threshold\cite{AB}. The s wave $\pi^{0}N$ elastic
scattering and charge exchange scattering length have been of considerable
interest since Weinberg\cite{W1} predicted that there should be an isospin
breaking effect due to the up, down quark mass difference. For
$a_{\pi^{0}N}$ this effect is $\simeq$ 30\%, in large part because the
isospin conserving term in $a_{\pi^{0}N}$ is very small ($\simeq 0.01
/m_{\pi}$) and consequently very difficult to measure. Since the charge
exchange scattering length is much larger ($\simeq 0.13/m_{\pi}$) this is
easier to observe. Here the isospin violating term due to the up and down
quark mass effect is the same (within a factor of $\sqrt{2}$) of the
elastic scattering prediction but in relative terms is $\simeq$ 2 to 3\% of
the isospin conserving term.  The most straightforward way to observe this
predicted isospin violation is to measure the cusp parameter $\beta$ in the
$\gamma p \rightarrow \pi^{0} p$ reaction with polarized proton
targets\cite{AB} and also $E_{0+}$ in the $\gamma p \rightarrow \pi^{+}n$
reaction. Then one could compare the values of $a^{cex}_{\pi^{+} n
  \rightarrow \pi^{0}p}$ with the measured value of $a^{cex}_{\pi^{-}p
  \rightarrow \pi^{0}n}$ from the line width in pionic hydrogen\cite{PSI}.
If isospin is a good quantum number then these will be equal and opposite.
Equivalently, the dynamic isospin breaking effect of the up, down quark
masses can be considered as an isospin breaking contribution to $\beta$ of
$\simeq$ 2 to 3\% \cite{AB}.

In addition to the s wave multipole $E_{0+}$ discussed above, ChPT makes
predictions for the threshold magnitudes of the three p wave multipoles
($P_{1}, P_{2},P_{3}$)\cite{BKM}. Since the p wave $\pi N$ phase shifts are
small at low energies, the p wave multipoles are essentially real.
Therefore for each threshold $\gamma N \rightarrow \pi N $ reaction there
are five multipole amplitudes to be measured ($Re(E_{0+}), Im(E_{0+}),
P_{1}, P_{2},P_{3}$). The unpolarized cross section can be written as
$\sigma(\theta)= A + B cos(\theta) + C cos^{2}(\theta)$ where A, B, and C
are real bilinear combinations of the five multipole
amplitudes\cite{BKM,formulas}. A complete experimental determination of the
multipoles requires then two additional polarization measurements. This
could include, e.g.  measurements with linear polarized photons and
unpolarized targets and with polarized targets and unpolarized photons
(this latter response measures imaginary parts of interference
amplitudes\cite{formulas}). 
Measurements with both photon and target polarization could also be
used\cite{formulas}. There are three independent isospin amplitudes and
four reaction channels ($\gamma p \rightarrow \pi^{+}n, \gamma p
\rightarrow \pi^{0}p, \gamma n \rightarrow \pi^{-}p, \gamma n \rightarrow
\pi^{0}n $).  so a measurement of all four constitutes a test of isospin
conservation. Therefore, a comprehensive set of measurements of the
threshold photo-pion reactions requires experiments with polarized photons
and targets, including both neutron and proton targets.
Unpolarized experiments on the threshold $\gamma p \rightarrow \pi^{0}p$
reaction have been performed \cite{Mainz,Sask}. The results for the s
wave multipole $E_{0+}$ were discussed above. In addition, two linear
combinations of the three p wave multipoles were found to be in agreement
with ChPT calculations \cite{BKM, Mainz,Sask}. More recently we have
performed a threshold $\vec{\gamma} p \rightarrow \pi^{0} p$ experiment at
Mainz with linearly polarized photons  and the data are presently
being analyzed. This will complete the measurement of the three p wave
multipoles for $\pi^{0}$ photoproduction from the proton.  However at the
present time, there are no measurements of $Im(E_{0+})$, which requires
polarized target experiments such as can be performed at SMASH.

There is also growing interest in using the deuteron as a neutron target
for the $\gamma n \rightarrow \pi N$ amplitudes. A ChPT calculation for the
coherent $\gamma D \rightarrow \pi^{0} D$ reaction exactly at threshold has
been performed\cite{Beane}, as was a first experiment of the $\gamma D
\rightarrow \pi^{0} X$ reaction (where X = D or np) at Saskatoon
\cite{Sask2}. Most $\pi^{0}$ spectrometers, including the one deployed at
Saskatoon, do not have sufficient energy resolution to determine whether
this reaction was coherent (X = D) or not (X=np). For the Saskatoon
experiment the smaller incoherent cross section was calculated with a model
and subtracted from the data to produce a coherent cross section to compare
with theory \cite{Sask2}. 

\begin{figure}[t]
  \begin{center}
    \epsfig{file=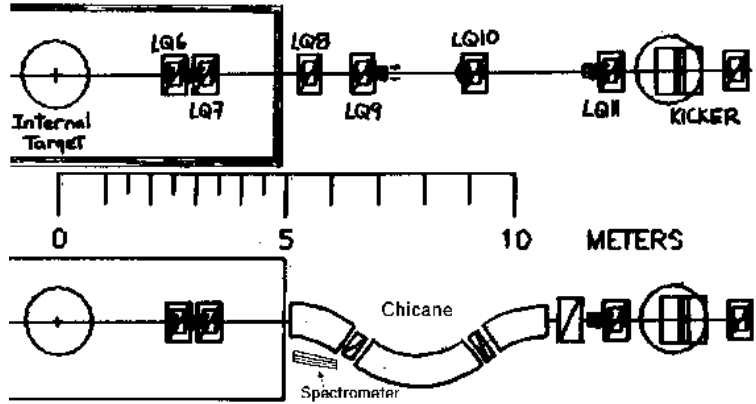,width=0.6\textwidth} \\
\hrulefill \\
    \epsfig{file=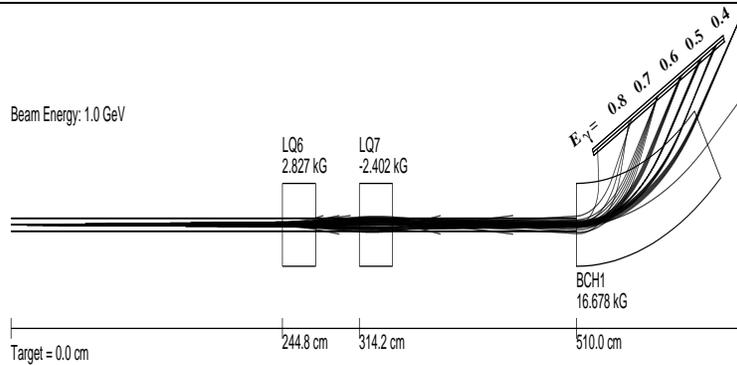,width=0.6\textwidth}   
    \caption{\footnotesize
       Conceptual layout for a very small angle electron
      tagging facility at the BLAST target position in the Bates storage
      ring. The top figure shows where the chicane
      would fit into the existing ring. The first dipole in the chicane
      would act as the electron spectrometer. The bottom figure shows
      sample scattered electron ray bundles (up to $\theta$=0.5$^{\circ}$ for 
      $\phi$=0,90,180$^{\circ}$). A
      schematicized wire chamber shows the corresponding almost--real photon
      energies that would be detected in this zeroth-order design.
      \normalsize}
    \label{fig:chicane}
    \label{fig:raytrace}
  \end{center}
\end{figure}

In general studies of the coherent $\gamma D \rightarrow \pi^{0} D$
reaction are best performed with recoil deuteron detection. This is an
important opportunity for polarized internal targets which are thin and
allow the detection of low energy recoil deuterons. There are also
opportunities in the $\Delta$ region. There an interesting sensitivity to
the quadrupole E2 amplitude has been theoretically demonstrated for
polarized deuteron targets \cite{WA} (see also Fig.\ref{fig:wilhelm}). 
This sensitivity is for small
$\pi^{0}$ CMS angles for which the recoil deuteron energy is small.

Although space does not permit a detailed discussion of Compton scattering,
a few remarks about its significance is in order. Previous measurements
have aimed to determine the electric and magnetic polarizability of the
proton. However, the spin dependent polarizabilities have yet to be
measured.  These spin dependent polarizabilities are subtle probes of the
internal structure of the nucleon. This requires Compton scattering of
polarized photons from polarized targets. The use of polarized internal
targets is an excellent way to make these measurements, particularly since
these targets are thin and one can measure the recoil nuclei as will be
discussed in the next section.
 
\section{\large SMASH:  {\it SM}all {\it A}ngle Electron {\it S}cattering  
  {\it H}odoscope \normalsize}

In this section we discuss the technique of very small angle electron
scattering, or almost--real photon tagging, with a specific conceptual
implementation for the Bates Storage 
Ring, and some ideas for important possible experiments. Much of this
material has 
been presented in an unpublished report from a previous incarnation of this
concept proposed at Bates several years ago \cite{Polite}. This previous
project was conceived before BLAST was funded, and therefore was presented
as a stand--alone facility with a different proposed position in the Bates
Ring. With the advent of BLAST funding, the new idea is to employ the BLAST
polarized target, and to make modifications to the ring to accommodate a very
small angle electron tagger.

The method of very small angle electron scattering has been known for many
years (see e.g. Ref.\cite{history}) and is based on the fact that the
virtual photons 
have very low $q^{2}$, so can be treated as almost--real photons.  
What is new about the present
proposal is the utilization of full polarization observables for both the
photon and the target, which is made possible by detecting these very small 
angle scattered electrons, thereby {\it tagging} the almost--real exchange
photons.The target polarization is made possible by the use
of internal targets. For the photon, one obtains circular polarization from
longitudinally polarized electrons.  The equivalent response function for
linear polarized almost--real photons \cite{formulas} are obtained by measuring
the $\phi$ dependence of the cross section\cite{Polite}. Therefore, this
virtual tagging proposal goes much further in utilizing the polarized,
pure, and windowless nature of internal targets, and the very large linear
polarization of the almost--real photons.

\subsection{Almost-real Photon Tagger}

A method to perform very small angle electron scattering/almost--real
photon tagging in the Bates Storage Ring is presented in Fig.
\ref{fig:chicane}. At this stage the design is conceptual only, but it does
exhibit the main features of what a fully designed facility would have. The
point here is to introduce the idea and its possibilities to motivate
further study.

The design is based on the introduction of a chicane downstream of the
internal target position, beginning just outside the BLAST superstructure.
A crucial element in the chicane design is that the extra path length
introduced must be $n\cdot\lambda=10.497cm$, where $\lambda$ is the beam
wavelength fixed by the RF.  In this conceptual design $n=5$. Another
important element is for the optics from the chicane to the next ring
dipole to be identical to the unmodified ring.  The chicane shown in
Fig.\ref{fig:chicane} does not satisfy this, but preliminary ring optics
investigations by T. Zwart of Bates \cite{zwart-pc} using box (not sector)
dipoles without the quadrupoles in between show that in principle this
should be possible with relatively minor adjustments of local beam line
elements.  A full study of this issue could not be completed in time for
this article, but nevertheless, one does not expect that the properties of
the fully designed chicane/tagger will be significantly different from what
is to be described.

\begin{figure}[ht]
  \begin{center}
    \begin{tabular}{l@{\hspace{0.60cm}}r}
      \epsfig{file=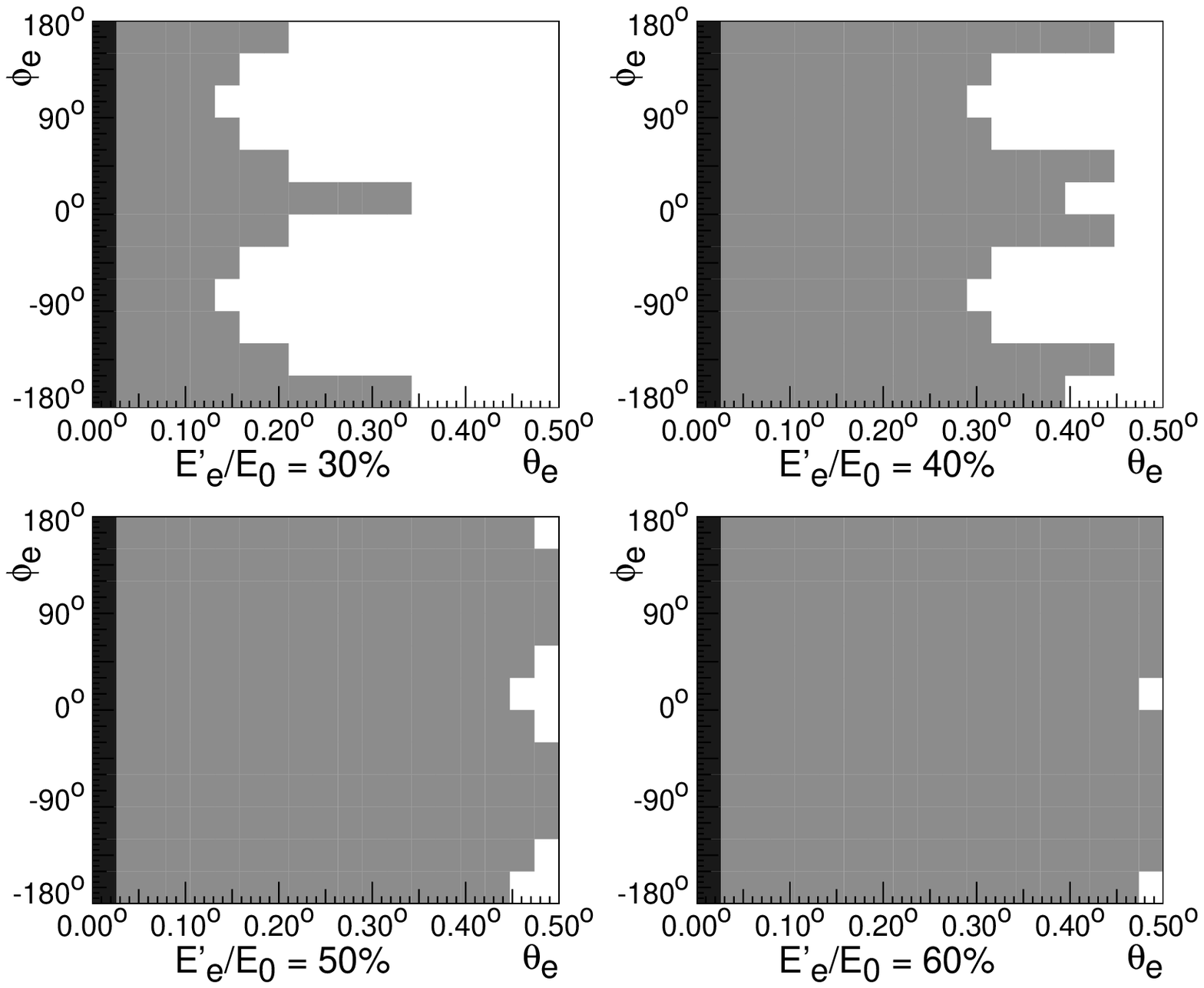,height=7.75cm,width=0.45\textwidth} &
      \epsfig{file=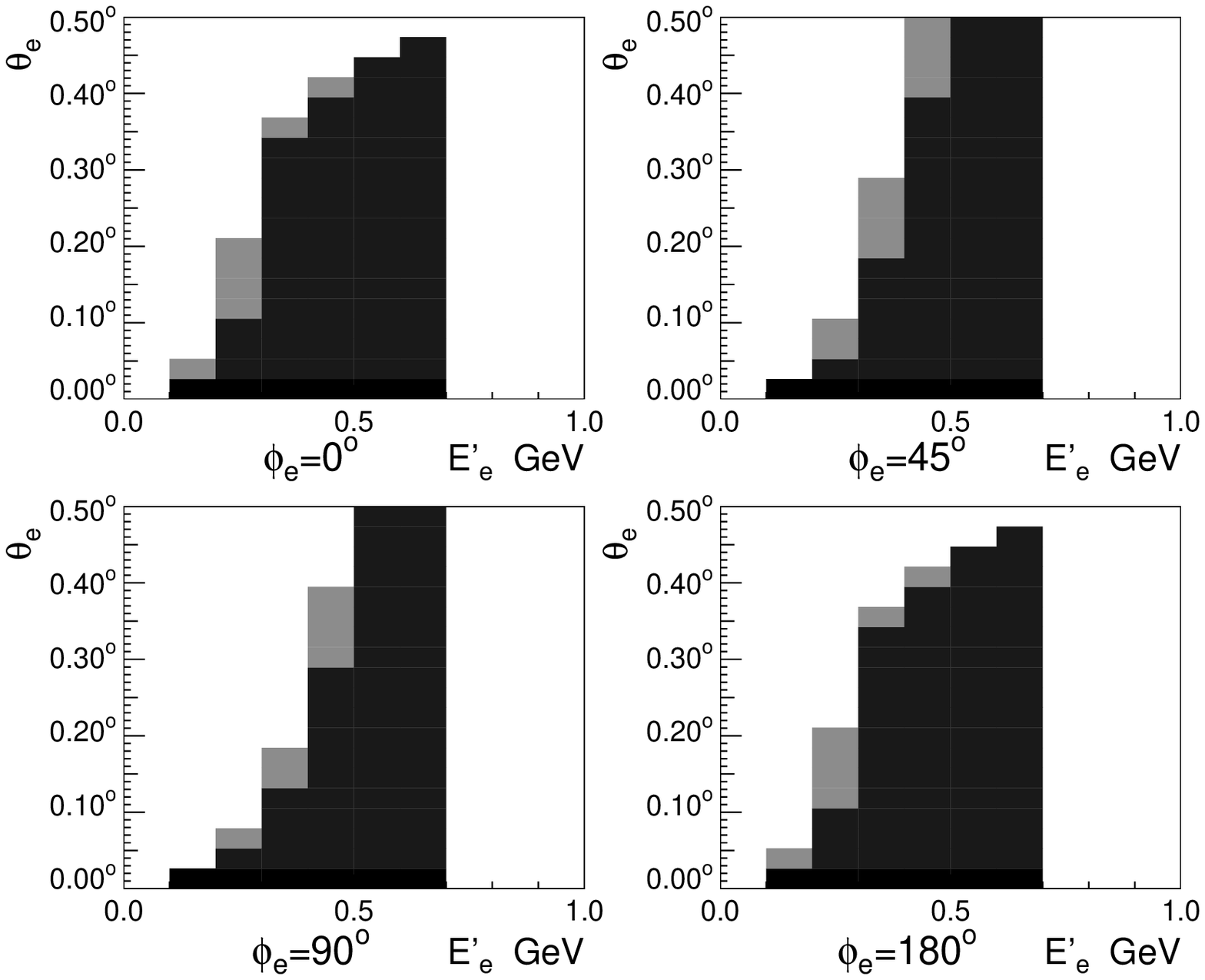,height=7.75cm,width=0.45\textwidth} \\
    \end{tabular}
    \caption{\footnotesize
Acceptance of very small angle scattered electrons for the
chicane spectrometer design in Fig.\protect\ref{fig:chicane}, 
{\bf Left:} $\phi_e$ $vs$ $\theta_e$ at fixed energy, and
{\bf Right:} $\theta_e$ $vs$ $E'$ at fixed azimuthal
angle.   The lightly shaded areas at right indicate partially filled bins.
     \normalsize }
    \label{fig:acceptance}
  \end{center}
\end{figure}

Beam electrons which do not interact in the target will be transported
around the chicane and then returned to the original ring trajectory.
However, the electrons that interact and are scattered into a very small
angle ($<0.5^{\circ}$ here) will have lost energy, so will be bent away
from the beam at the first chicane dipole (see Fig.\ref{fig:raytrace}). By
placing wire chambers to detect these scattered electrons outside this
dipole, a QQD spectrometer will be realized. A ray tracing
simulation shown in Fig.\ref{fig:raytrace} shows that something
approximating a focal plane emerges from even this simple design.  One is
confident then that with a carefully designed first chicane dipole a
spectrometer with reasonable optical properties can be achieved.

As a guideline to what can be expected in a full--fledged system, the
acceptance of the chicane system (Fig.\ref{fig:raytrace}) is shown in
Fig.\ref{fig:acceptance} as a function of the ratio of the outgoing to
incident electron energy E'/E, and the outgoing electron angles
$\theta_{e}, \phi_{e}$. The acceptance is limited by the apertures of the
first two quadrupoles. In this case a rectangular beam box was used, which
increases the acceptance at $\phi= 0, 90, 180, 270^{\circ}$. With the usual
cylindrical beam pipe, the acceptance is limited to about 0.33$^{\circ}$.
Clearly, larger aperture quadrupoles would be preferable, but nonetheless
even in this scenario decent momentum coverage is achieved, and as will be
shown, a $\simeq0.5^{\circ}$ angular acceptance is not unreasonable for
almost--real photon tagging.

Here we recall the formulas for the kinematic variables relevant to small
angle electron scattering.  Note that in the extreme forward direction, the
finite mass of the electron cannot be neglected, so the exact expressions
must be used.  These are \cite{Polite}:
\begin{eqnarray*}
q^2   ~=~-Q^2& = & 2m_e^2 -2EE'+2pp'cos\theta_e\\
k_\gamma & = & \sqrt{p^2+p'^2-2pp'cos\theta_e} \\
\epsilon & = & (1+\frac{Q^2 |k_\gamma|^2}{2p^2p'^2sin^2\theta_e})^{-1}\\
P_{\gamma} & = & h\cdot\sqrt{1-\epsilon^{2}} \\
\Gamma   & = &  \frac{\alpha}{2\pi^2}
                \frac{E'}{E}
                \frac{k_\gamma}{Q^2}
                \frac{1}{1-\epsilon}
\label{eqn:formulas}
\end{eqnarray*}
where $\theta_e$ is the electron scattering angle, $E (E')$ and $p (p')$
are the beam energy and momentum of the incident (scattered) electrons, $h$
is the longitudinal beam polarization, and $q^2$, $k_\gamma$, $\epsilon$,
$P_{\gamma}$, and $\Gamma$ are the virtual photon four-momentum, momentum,
transverse polarization, circular polarization, and flux, respectively.
These formulas still neglect radiative corrections, but these
  should be relatively small, at the 1\% level.

\begin{figure}[ht]
  \begin{center}
    \mbox{\epsfig{file=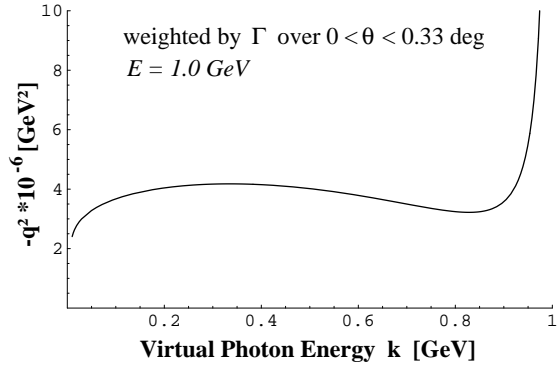,width=0.45\textwidth}}
    \caption{\footnotesize
      Four-momentum transfer of very small angle scattered 1 GeV electrons
      versus virtual photon energy, averaged over $\theta <
      \frac{1}{3}^{\circ}$ and weighted by the virtual photon flux
      $\Gamma$. Note that these virtual
      photons are essentially real, except those in the untaggable region
      very near the endpoint.  
\normalsize}
    \label{fig:avgQ2}
  \end{center}
\end{figure}

Figure \ref{fig:avgQ2} shows the four--momentum transfer of the detected
scattered electrons, averaged over a $0<\theta_{e}<0.33^{o}$ acceptance and
weighted by the virtual photon flux.  The very small values indicate that
 the exchanged virtual photons are essentially real, except very near the
 endpoint, which nevertheless fall outside the
tagging region.

\begin{figure}[ht]
  \begin{center}
    \begin{tabular}{l@{\hspace{1cm}}r}
      \epsfig{file=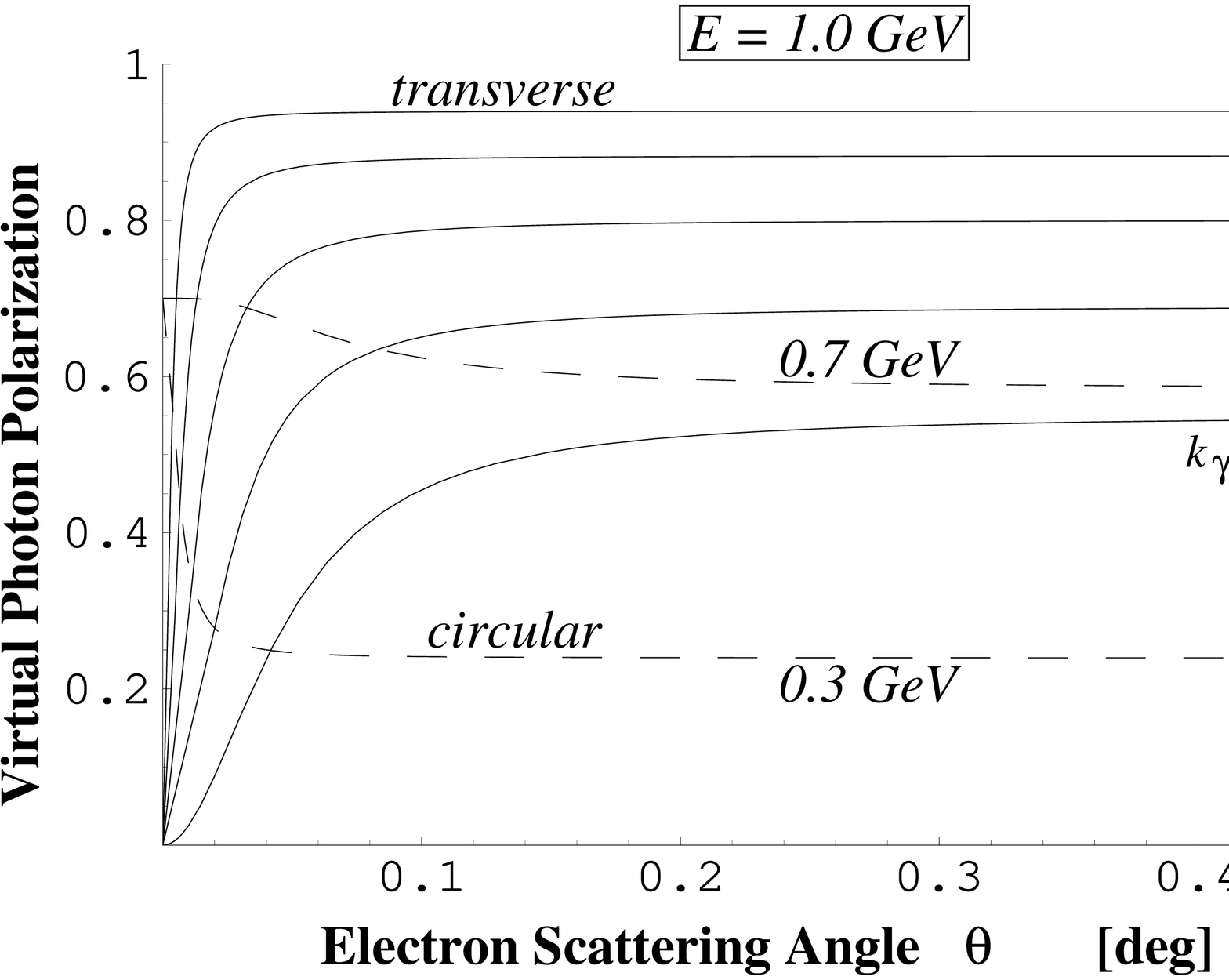,width=0.45\textwidth} &
      \epsfig{file=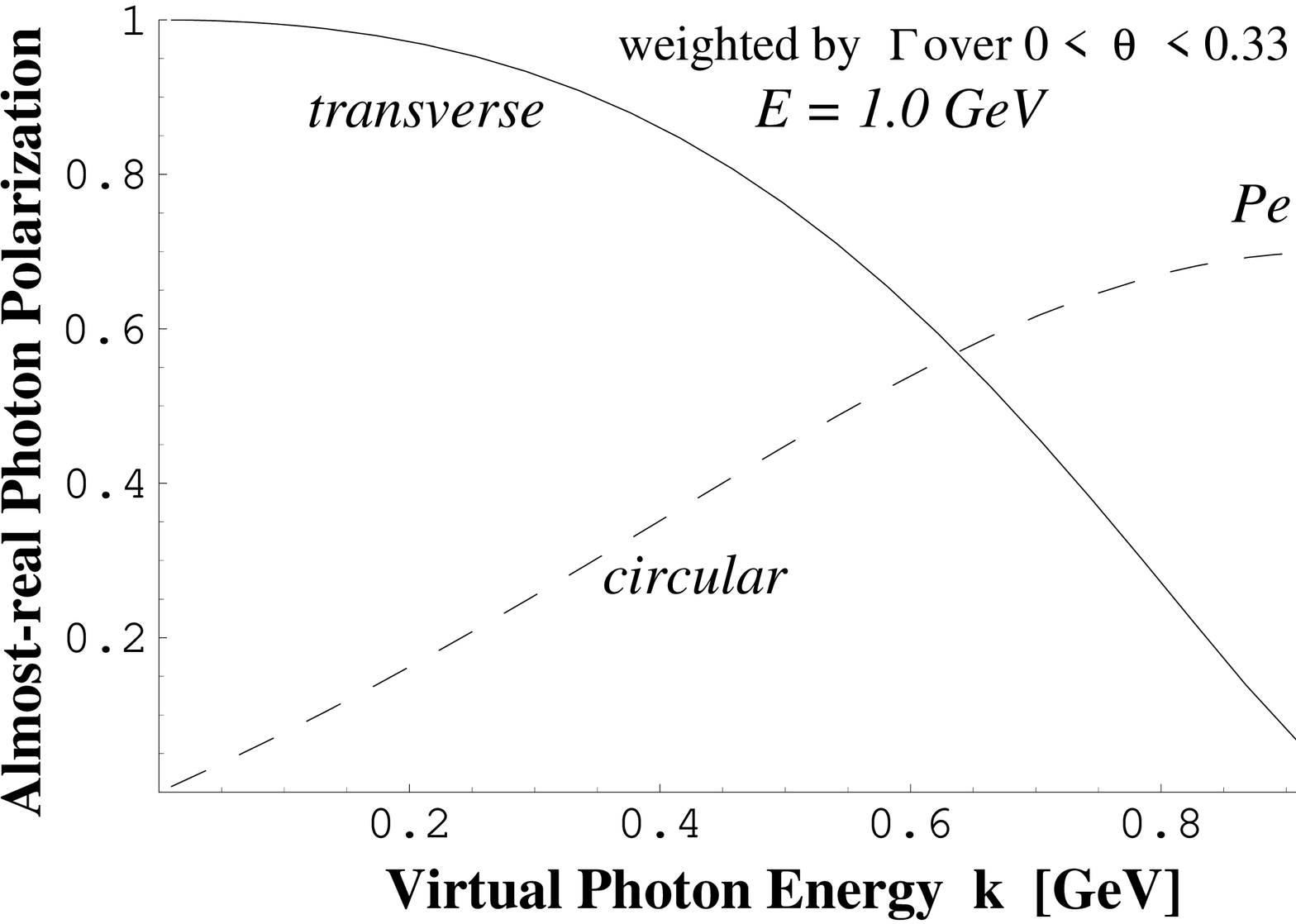,width=0.45\textwidth} \\
    \end{tabular}
    \caption{\footnotesize
      {\bf Left:} Virtual photon polarizations versus electron scattering
      angle for 1 GeV electrons for various values of photon energy. ; 
      {\bf Right:} Virtual photon polarization versus photon energy,
      averaged over $\theta < \frac{1}{3}^{\circ}$ electron scattering
      angle and weighted by the virtual photon flux $\Gamma$. The solid
      (dashed) line shows the transverse (circular) polarization.
      \normalsize}
    \label{fig:epsilon}
  \end{center}
\end{figure}

The transverse polarization of almost--real photons are shown
in Fig.\ref{fig:epsilon} both as a function of scattering angle at fixed
photon energy, and also versus energy averaged over angle weighted by the
virtual photon flux.  Note the very high transverse polarizations over the
entire tagging range, averaging about 70\%. Note as well that the
polarization is constant after about 0.1$^{o}$.  Also shown in
Fig.\ref{fig:epsilon} is the transfered circular polarization, assuming a
70\% longitudinally polarized electron beam. Sizable polarizations which
are very flat with angle are seen here as well.

\begin{figure}[ht]
  \begin{center}
    \begin{tabular}{l@{\hspace{1cm}}r}
      \epsfig{file=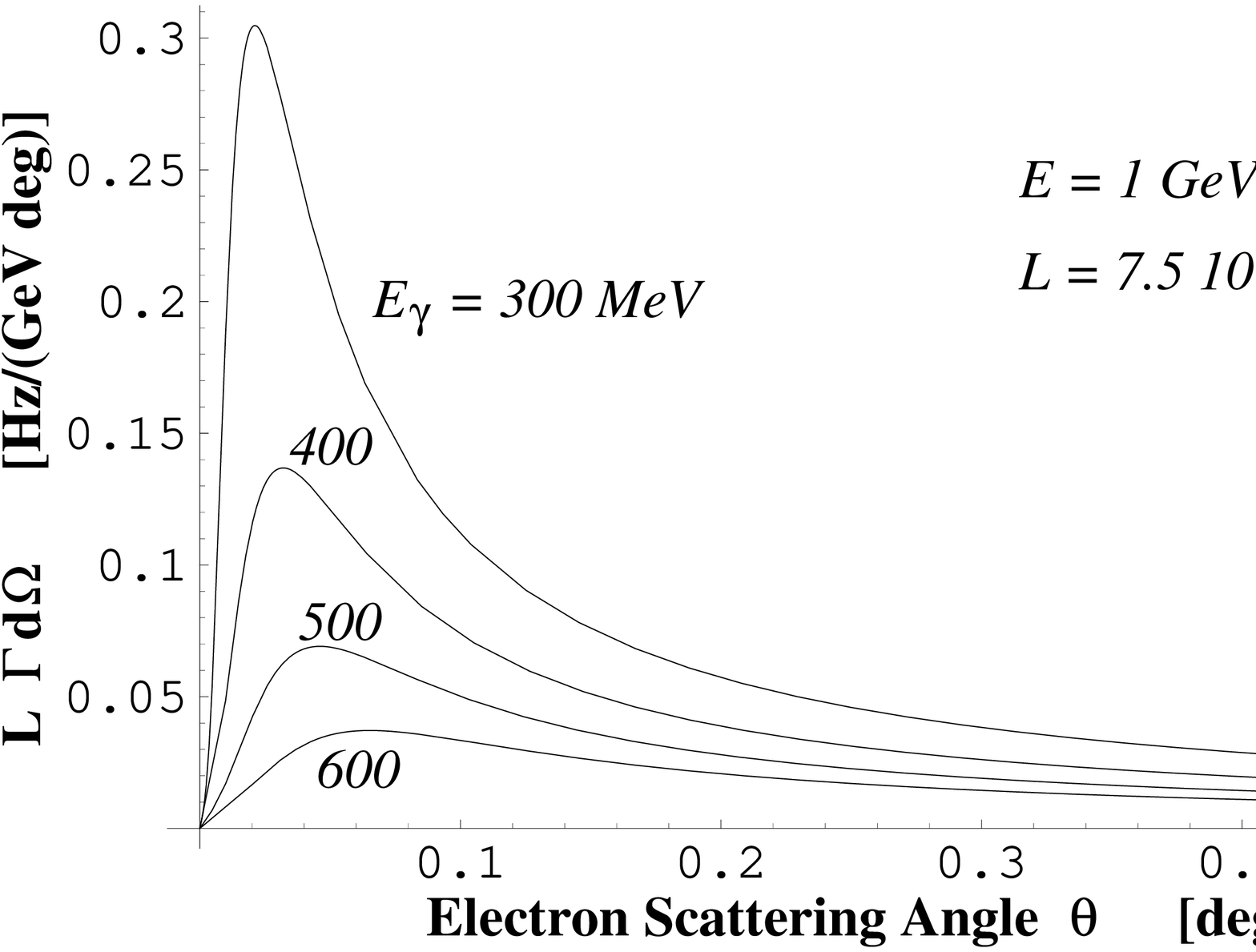,width=0.4\textwidth} &
      \epsfig{file=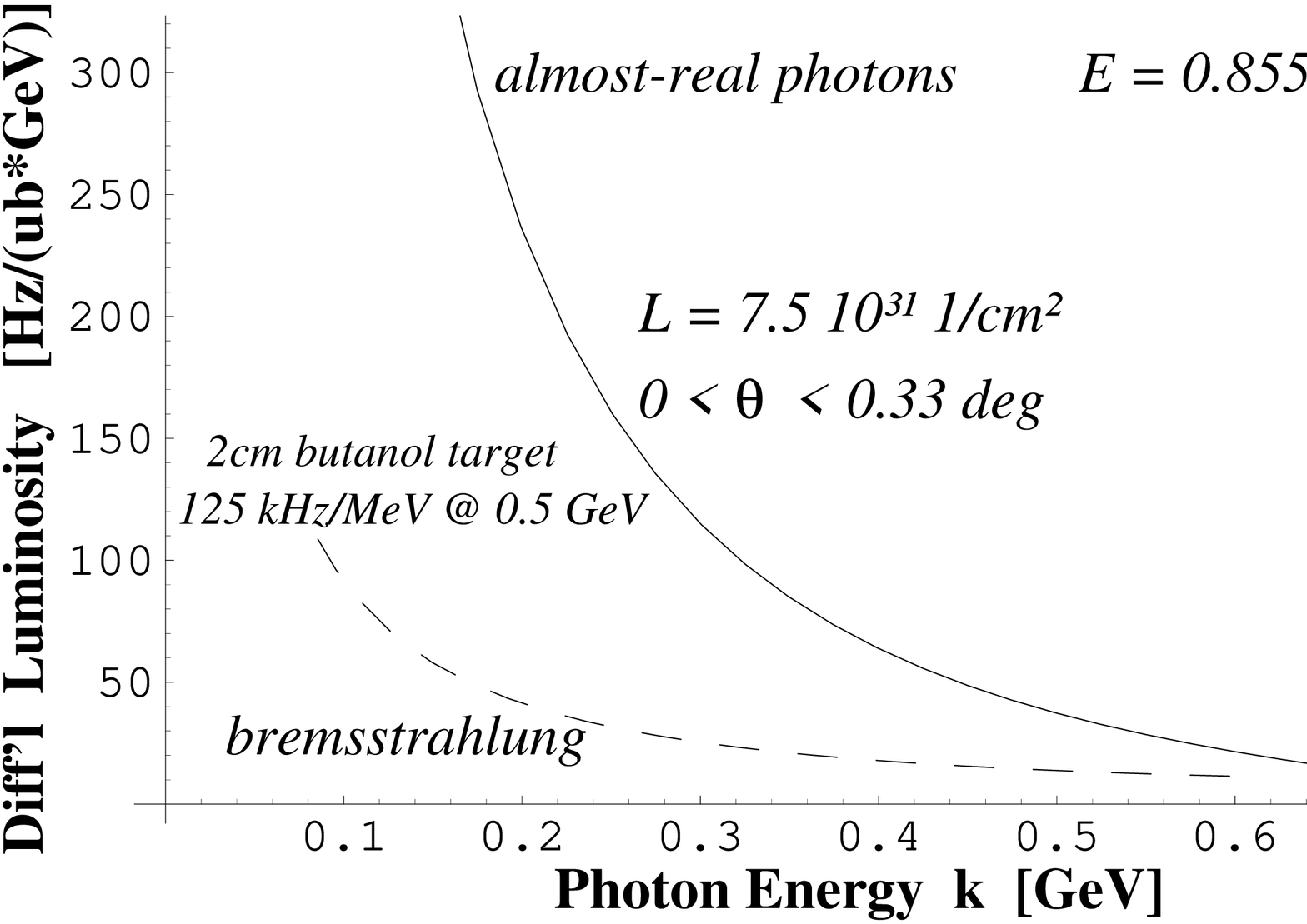,width=0.4\textwidth} \\
    \end{tabular}
    \caption{\footnotesize
      {\bf Left:} Virtual photon rate versus electron
      scattering angle, for a 1 GeV beam of luminosity 7.5
      $\cdot$10$^{31}$ expected for the BLAST internal polarized proton target.
      {\bf Right:} Virtual photon flux integrated over
      $0<\theta<$0.33$^{\circ}$ solid angle versus virtual photon energy,
      for the same luminosity at 0.855 GeV. Also shown is the
      bremsstrahlung rate at that energy expected from the Mainz tagged
      photon facility, assuming a 2cm polarized butanol target.
            \normalsize}
    \label{fig:virtphotflux}
  \end{center}
\end{figure}

The virtual photon flux is shown in Fig.\ref{fig:virtphotflux}, multiplied
by the luminosity 7.5$\cdot$10$^{31}$ cm$^{-2}$, which is what is expected
at BLAST for the internal polarized proton target. The left figure shows
the flux versus angle, where it is seen that the flux is strongly forward
peaked at most energies, and that there is diminishing strength beyond
about 0.5$^{o}$. Although a larger angular acceptance is clearly
preferable, especially for the highest photon energies, this shows that the
returns are diminishing beyond this range.  The total flux versus photon
energy is also shown at right in Fig.\ref{fig:virtphotflux}. 

\begin{figure}[ht]
  \begin{center}
      \epsfig{file=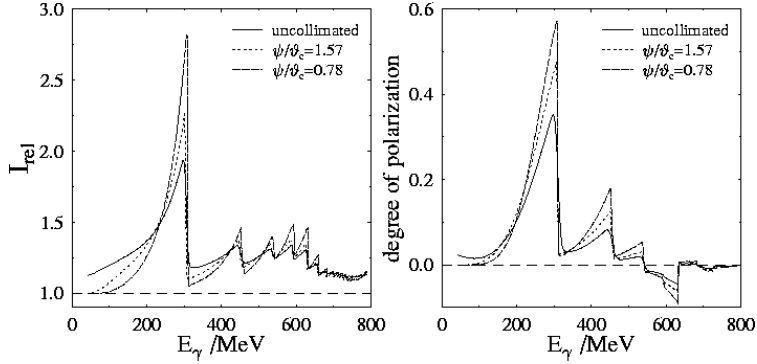,width=0.6\textwidth} 
    \caption{\footnotesize
{\bf Left:} Ratio of coherent bremsstrahlung spectrum from a diamond radiator to the
incoherent spectrum at E$_{0}$=0.855 GeV, and {\bf Right:} degree of linear 
polarization, as a function of photon energy, at the Mainz tagged photon
facility (see Ref. \protect\cite{mami-a2}). Note that here the linear
polarization is sizable in only a narrow energy range, unlike that shown in 
Fig.\protect\ref{fig:epsilon}.
      \normalsize}
    \label{fig:coh_brem}
  \end{center}
\end{figure}

The incoherent bremsstrahlung flux assuming a 2$cm$ butanol target from a
modern tagged photon facility at Mainz \cite{mami-a2} is shown as the dotted line in
Fig.\ref{fig:virtphotflux}.  As well, the energy and linear polarization
spectrum of coherent bremsstrahlung beam using a diamond radiator  is shown
in  Fig.\ref{fig:coh_brem}.  The former figure demonstrates that even with a thin
internal target, a sizable rate advantage is seen using tagged almost--real 
photons, especially for $\frac{E_{\gamma}}{E_{0}} < 0.5$.  Moreover, low
energy recoils are not accessible with usual (thick) frozen targets 
like butanol,
whereas they are with (thin) internal targets. The coherent bremsstrahlung
linear 
polarization spectrum (Fig.\ref{fig:coh_brem}) shows
lower polarizations over a much smaller energy range than 
those in almost--real photon tagging (Fig.\ref{fig:epsilon}).

The afore described figures demonstrate the salient features of a very small 
angle electron spectrometer of the kind shown in
Fig.\ref{fig:chicane}. Namely, it tags virtual photons which are
essentially "real" with high transverse and circular polarizations, and a
large flux. Coupled with thin highly polarized internal targets, and a
versatile detector such as BLAST, supplemented with a hodoscope for low
energy recoils (such as a large coverage Si strip counter), 
it is clear that an almost--real photon
tagger opens up many new opportunities. Some examples will be described in the
following section.

\subsection{Example Experiments}

Here a few example experiments will be discussed, focusing on those
previously mentioned in conjunction with the chiral dynamics studies
outlined in the introduction.  Note that the experimental details have not
been worked out in time for this contribution, so only a broad outline of
what is required will be offered to motivate future study.

\begin{figure}[ht]
  \begin{center}
      \epsfig{file=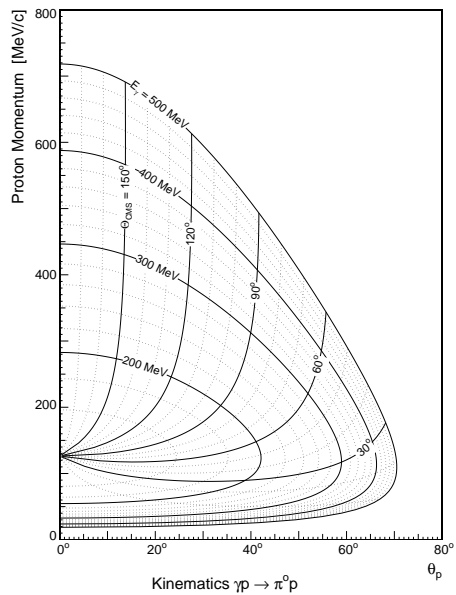,width=0.35\textwidth} 
    \caption{\footnotesize
Proton momentum versus scattering angle in neutral pion photoproduction
near threshold, with contours of constant photon energy and pion CMS
angle. Note that close to threshold the protons recoil forward $<20^{\circ}$.
      \normalsize}
    \label{fig:pi0photokin}
  \end{center}
\end{figure}

Figure \ref{fig:pi0photokin} shows the recoil proton momentum versus lab
scattering angle in the $\gamma p \rightarrow \pi^{0}p$ reaction for
constant photon energy and constant CMS pion scattering angle. One observes 
that near threshold the protons recoil at low energy in a forward cone, so
that to detect these a forward detector must be constructed. Given the
constraints of the BLAST internal target, this detector would probably need
to be compact, therefore of high positional resolution, implying a silicon
strip--type unit of the kind used in many high energy experiments. Note
that thin windowless internal targets have the great advantage here of
allowing the low momentum recoils to be detected with minimal interference.  
One way to identify the scattered $\pi^{0}$s is to detect the recoil
protons with sufficient energy and angular resolution to determine the 
missing mass. Another is to detect the produced $\pi^{0}$s, and for this 
a crystal "ball" or "cylinder" could be 
constructed to fit around the target. This would greatly reduce backgrounds 
and add flexibility. It would probably necessitate
operating with the BLAST detector "pulled apart", or else BLAST could
remain in place and photon detectors installed outside the magnet, but this 
option increases the size and cost, and reduces the solid angle.

The same Fig.\ref{fig:pi0photokin} can be used to show the kinematics for
the $\gamma p \rightarrow \pi^{+}n$ reaction, although due to the final
state mass differences, the threshold and kinematic contours will be
slightly altered.  In this case one could use the BLAST detector for
$\pi^{+}$ detection with good angular coverage.  The planned neutron
detector would cover $38<\theta_{n}<70^{0}$, which would cover nicely the
$\Delta$ resonance region for e.g. $\gamma p \rightarrow \Delta$ studies.
However, this precludes the near threshold region, so these detectors would 
need to be shifted somewhat, or new detectors constructed, to cover the
more forward angles.

\begin{figure}[ht]
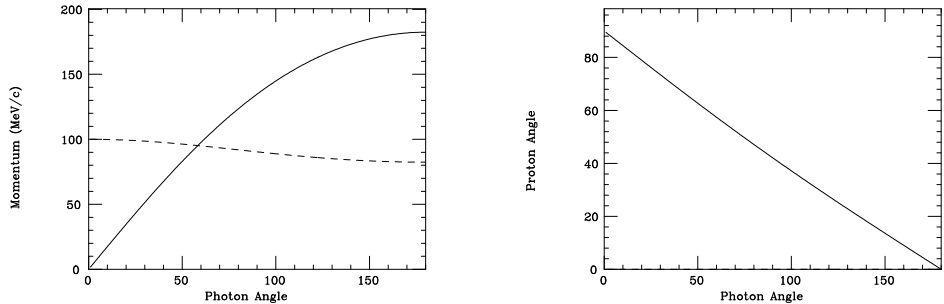

  \begin{center}
    \begin{tabular}{l@{\hspace{0.25cm}}r}
      \epsfig{file=compfig7.eps,width=0.3\textwidth, angle=90} &
      \epsfig{file=compfig8.eps,width=0.3\textwidth, angle=90} \\
    \end{tabular}
    \caption{\footnotesize
{\bf Left:} Proton and photon momentum versus photon lab angle for
Compton scattering at E$_{\gamma}=100 MeV$, and   {\bf Right:} Angular
correlation of the scattered photon and recoil proton.
      \normalsize}
    \label{fig:compton}
  \end{center}
\end{figure}

The proton and photon kinematic relationships for Compton scattering at
E$_{\gamma}=100 MeV$ are shown in Fig.\ref{fig:compton}.  This experiment
would use the same or very similar setup to that used for the near
threshold $\gamma p \rightarrow \pi^{0}p$ experiment described above to
detect the scattered photon and recoil proton. In addition, near forward 
photon angles are of interest in double--polarized Compton scattering (see
reference \cite{Polite}). Here, the proton momentum is low and the angle
large, and so can be detected with the proposed BLAST low energy recoil
detector \cite{Blast}. Again, in both cases the thin windowless internal
target is seen to be greatly advantageous to facilitate low energy proton
detection.  Circularly polarized photons are required, and we have seen
(Fig.\ref{fig:epsilon}) that these can be rather sizable.

\begin{figure}[ht]
  \begin{center}
      \epsfig{file=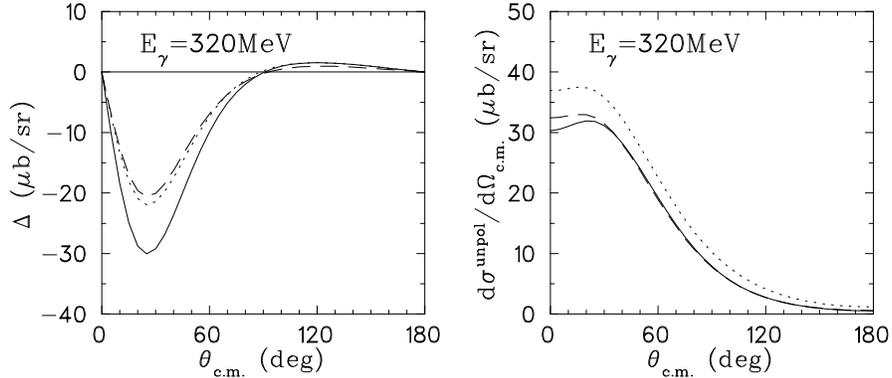,width=0.3\textwidth, angle=90}
    \caption{\footnotesize
{\bf Left:} Cross section difference ($\sigma\uparrow - \sigma\downarrow$)
 and {\bf Right:} cross section for coherent $\pi^{0}$
photoproduction on a $\pm$100\% vector polarized deuteron target at the $\Delta$ 
energy, from the model described in Ref.\protect\cite{WA}. The solid line
is the full calculation, whereas the dashed line has the quadrupole E2
amplitude removed.   The region of largest sensitivity corresponds to large 
angle recoil deuterons of a few MeV kinetic energy.
      \normalsize}
    \label{fig:wilhelm}
  \end{center}
\end{figure}

For a final example, consider the coherent $\gamma D \rightarrow
\pi^{0} D$ reaction on a vector polarized deuteron target 
which has been shown (see \cite{Polite}) to be
sensitive to the  nucleon quadrupole E2 amplitude.  Figure
\ref{fig:wilhelm} demonstrates the calculated sensitivity, which 
is greatest at small pion CMS angles where the recoil deuteron energy is
low. Again the proposed BLAST low energy recoil detector would be used
to detect these deuterons.  The neutral pions would be detected with the
same setup used for the threshold photoproduction and Compton scattering
experiments. Once again the merits of a thin windowless internal target is
seen.

\section{Conclusion}

The experiments presented in the previous section are but a few examples of
what can done to 
exploit the unique capability of an almost-real photon tagger coupled with 
the proposed BLAST facility \cite{Blast}. Including also
a new forward hadron detector and a large acceptance photon detector would
not only allow three important Chiral Dynamics experiments to be
done, but should 
also open up a whole new arena of unique experiments. 
The low energy QCD experiments include : threshold photo-$\pi^{0}$ 
production on polarized protons to measure $Im (E_{0+})$ which is 
sensitive to the isospin breaking due to the light quark mass 
differences, fully polarized Compton scattering, which measures the 
internal  quark helicity structure, and the electric quadrupole 
contribution to the $\gamma N \rightarrow \Delta$ transitions in the 
proton and the deuteron. Other reactions not touched on here include 
photo nucleon and photo pion production from polarized few body 
nuclei. We believe that this opens a new and exciting window of 
opportunity for polarized, internal, targets at  Bates.
More detailed design efforts are currently underway, and more input from
the collaboration in general would be warmly welcomed.


\end{document}